# Improving News Recommendations through Hybrid Sentiment Modelling and Reinforcement Learning

Eunice Kingenga, u13268831@tuks.co.za
Mike Wa Nkongolo, mike.wankongolo@up.ac.za
Department of Informatics, University of Pretoria

**Abstract.** News recommendation systems rely on automated sentiment analysis to personalise content and enhance user engagement. Conventional approaches, however, often struggle with ambiguity, lexicon inconsistencies, and limited contextual understanding, particularly in multi-source news environments. Existing models typically treat sentiment as a secondary feature, reducing their ability to adapt to users' affective preferences. To address these limitations, this study develops an adaptive, sentiment-aware news recommendation framework by integrating hybrid sentiment analysis with reinforcement learning. Using the BBC News dataset, a hybrid sentiment model combines VADER, AFINN, TextBlob, and SentiWordNet scores to generate robust article-level sentiment estimates. Articles are categorised as positive, negative, or neutral, and these sentiment states are embedded within a Q-learning architecture to guide the agent in learning optimal recommendation policies. The proposed system effectively identifies and recommends articles with aligned emotional profiles while continuously improving personalisation through iterative Q-learning updates. The results demonstrate that coupling hybrid sentiment modelling with reinforcement learning provides a feasible, interpretable, and adaptive approach for user-centred news recommendation.

**Keywords:** news recommendation systems, sentiment analysis, reinforcement learning, information retrieval

## I. Introduction

### A. Background and Motivation

The proliferation of digital news platforms has fundamentally changed how individuals consume information. Users are increasingly exposed to vast amounts of content, making the task of identifying relevant and engaging articles more challenging. Traditional recommendation systems, such as content-based filtering and collaborative filtering, focus primarily on explicit user interactions like clicks, ratings, and browsing histories. While these approaches have achieved measurable success in personalizing content, they often fail to account for the affective dimension of news consumption, which can significantly influence user engagement and satisfaction [1], [2]. Sentiment analysis provides a mechanism to capture the emotional tone of textual content. By quantifying the polarity of articles, it allows recommendation systems to align suggestions with users' emotional preferences. However, conventional sentiment analysis methods, particularly those relying on a single lexicon or simple machine learning classifiers, face several challenges. They often struggle with ambiguity, domain-specific nuances, and lexicon inconsistencies, especially in multi-source news environments [3], [4]. Moreover, existing recommendation systems frequently treat sentiment as an auxiliary feature rather than as an integral component influencing the recommendation policy, limiting their adaptability to user-specific affective dynamics.

## B. Problem Statement

Despite advances in recommendation algorithms, several critical limitations remain in current approaches:

1. **Weak Sentiment Integration:** Conventional systems rarely incorporate sentiment as a state feature for adaptive recommendation policies, limiting personalization accuracy.

2. **Contextual Ambiguity:** Lexicon-based sentiment classifiers often fail to capture nuanced meanings, while machine learning models can underperform without sufficient context or domain-specific training data.

3. **Limited Adaptivity:** Static recommendation models cannot dynamically adjust to evolving user preferences over time, particularly when user behavior is influenced by emotional factors.

These limitations result in reduced recommendation accuracy, poor alignment with user sentiment, and limited effectiveness in reinforcement-driven personalization systems.

## C. Prior Work

Existing research in news recommendation has explored various strategies, including:

- **Content-Based Filtering:** Leveraging article metadata and textual features to match users with relevant content [1].

- **Collaborative Filtering:** Utilizing historical user behavior patterns to predict preferences [2].

- **Sentiment-Aware Recommenders:** Incorporating sentiment analysis as an auxiliary feature to improve engagement [3].

However, these approaches either underutilize sentiment information or fail to combine sentiment modeling with adaptive reinforcement learning, limiting their ability to dynamically optimize recommendations based on emotional alignment.

## D. Contribution of This Study

To address the aforementioned challenges, this study proposes a novel, hybrid approach that integrates Exploratory Data Analysis (EDA), multi-lexicon sentiment modeling, and reinforcement learning for news recommendation. Key contributions include:

1. **Hybrid Sentiment Modeling:** Combining VADER, AFINN, TextBlob polarity, and SentiWordNet scores to produce robust article-level sentiment estimates.

2. **Sentiment-Aware Reinforcement Learning:** Embedding sentiment categories as state representations in a Q-learning framework to optimize recommendation policies based on sentiment-aligned reward signals.

3. **Adaptive Personalization:** The system continuously improves user-centered recommendations through iterative Q-learning updates, enhancing both engagement and alignment with user affective preferences.

This integration demonstrates the practical potential of coupling hybrid sentiment analysis with reinforcement learning, overcoming limitations of prior methods and advancing adaptive, sentiment-driven news recommendation systems. The remainder of this paper is organized as follows: Section II presents related work in sentiment-aware news recommendation and reinforcement learning-based recommender systems. Section III details the proposed hybrid sentiment analysis and reinforcement learning framework. Section IV presents the results, followed by a discussion in Section V. Finally, Section VI concludes the paper and outlines future research directions.

## II. Related Work

### A. Sentiment-Aware and Affective News Recommendation

The role of sentiment or emotional tone in news recommendation has garnered increasing attention, as users' engagement and satisfaction often depend not only on topical relevance but also on the affective character of content [4]. Early work in sentiment-aware recommendation focuses on ensuring diversity in sentiment orientation among recommended articles to avoid reinforcing sentiment "echo chambers." For example, *SentiRec*: Sentiment Diversity-aware Neural News Recommendation implements a sentiment-aware news encoder, jointly training a sentiment-prediction auxiliary task along with a click-ranking objective, and introduces a regularization to encourage sentiment diversity in the recommendation list. Experiments demonstrate that the method increases sentiment variety without sacrificing recommendation performance [5]. More broadly, the literature has begun analyzing sentiment and stance biases in news recommender systems. The study by Alam et al. [2] applies sentiment and stance detection to a corpus of news on migration and demonstrates a tendency of several recommenders to favor negative-sentiment articles and stances aligned with preexisting user biases, highlighting that sentiment-aware recommendation must also consider fairness and bias amplification risks. Hybrid sentiment-classification + recommendation designs have also been proposed outside purely neural-network models. For example, a recent work by Xi and Jiang [6] combines multiple sentiment classifiers and text-similarity recommendation logic to improve sentiment classification precision (reported at 0.95) and enhance recommendation quality, suggesting that hybrid sentiment models can support effective news recommendation frameworks. Nonetheless, in many of these studies, sentiment remains a *secondary* feature or a regularization term — the recommendation mechanism itself often remains static, lacking adaptivity to changing user affective states or sentiment dynamics over time. This limits their effectiveness in scenarios where user preferences evolve or where the emotional tone of news content matters for longer-term engagement.

## B. Deep / Knowledge-Aware News Recommendation (Non-RL)

In parallel, the field has explored deep learning approaches to news recommendations that focus on richer content semantics and external knowledge, rather than sentiment. For instance, DKN: Deep Knowledge-Aware Network for News Recommendation integrates knowledge-graph entities with word-level representations via a word-entity aligned convolutional neural network, plus an attention-based user history aggregation module. This allows the system to exploit semantic and entity-level common-sense knowledge, leading to improved click-through predictions compared to purely text-based models [7]. Other hybrid and session-based news recommenders such as a recurrent neural network–based system presented by Gabriel De Souza et al. [8] incorporate additional information like article recency, popularity, and session context to model short-term user interests and improve recommendation accuracy and novelty over baseline item-based approaches. Such deep and knowledge-aware methods improve representational richness and recommendation relevance, yet typically do not incorporate sentiment or affective signals; meaning they may overlook the emotional dimension that influences how users respond to news.

## C. Reinforcement Learning With Adaptive and Dynamic News Recommendation

Recognizing that news consumption is dynamic with rapidly changing content, evolving user interests, and the temporal value of engagement, several works cast news recommendation as a sequential decision problem, solvable via reinforcement learning (RL). Notably, *DRN: A Deep Reinforcement Learning Framework for News Recommendation* [9] formulates news recommendation as a Markov Decision Process and applies Deep Q-Learning to optimize not only immediate reward (e.g., click-through) but also future reward (e.g., long-term user engagement, and return frequency). The authors argue that RL helps avoid recommendation stagnation (monotony) and better handle rapid news turnover and evolving user preferences. More broadly, recent surveys and reviews of deep recommendation systems highlight RL-based methods as a promising approach for long-term personalized recommendation and dynamic user modeling [10]. Complementary recent work seeks to address noise in implicit feedback (common in news recommendation: clicks, dwell time, and skips). For example, the study *Denoising Neural Network for News Recommendation with Positive and Negative Implicit Feedback* incorporates both positive and negative implicit feedback (e.g., clicked vs. seen-but-not-clicked) to denoise user feedback signals, improving recommendation robustness and overall performance [11]. These adaptive and denoising approaches demonstrate that dynamic, sequential, feedback-aware recommendation systems can handle the inherent challenges of real-world news platforms like content volatility, sparse and noisy feedback, and changing user interests.

## D. Summary: Gaps and Opportunity

From the review above, two largely parallel strands emerge in the literature: *(1) sentiment-aware recommenders (focusing on emotional tone, diversity, and bias), but typically static or non-sequential; (2) adaptive, dynamic recommendation via deep /*

*RL-based models, but typically focused on content relevance, recency and click behavior, without explicit modeling of sentiment.* What remains relatively unexplored and constitutes the opportunity our work addresses is *the integration of robust sentiment modeling (e.g., via hybrid sentiment analysis) as a first-class, and dynamic state signal within a reinforcement-learning news recommender*. Embedding sentiment as part of the user or content-state enables a recommender to adapt not only to changing topics or interests but also to evolving emotional preferences or mood, potentially improving both engagement and satisfaction in a more human-aware manner.

## III. Methodology

The workflow of the methodological pipeline employed to construct the proposed sentiment-aware reinforcement learning (RL) news recommendation framework is illustrated in Figure 1. The methodology integrates data preprocessing, hybrid lexicon-based sentiment modelling, probabilistic sentiment encoding, and Q-learning-driven policy optimisation [12].

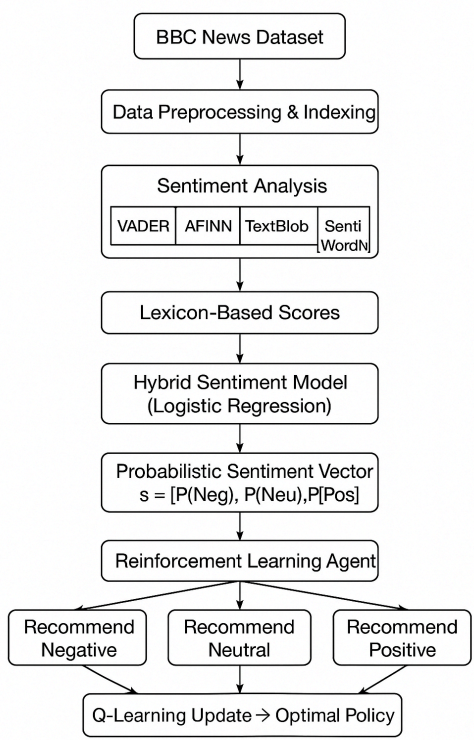

**Figure 1.** Proposed computational pipeline

**Dataset Acquisition and Indexing.** The study uses the BBC News dataset [4], which provides curated, and multi-category news articles suitable for text analytics and recommendation research (https://www.kaggle.com/code/gpreda/bbc-news-rss-feeds). Each article is assigned a unique index and metadata fields necessary for downstream modelling, including title, body text, and category labels (Figure 2).

**Data Preprocessing.** To ensure high-quality sentiment estimation, the textual data undergo standard natural-language preprocessing [13]. The steps include: Tokenization of article text

66

into semantic units, lowercasing to maintain lexical consistency, stop-word removal to reduce non-informative tokens, punctuation stripping and whitespace normalization, and lemmatization, ensuring that tokens map to their canonical morphological forms [4], [13].

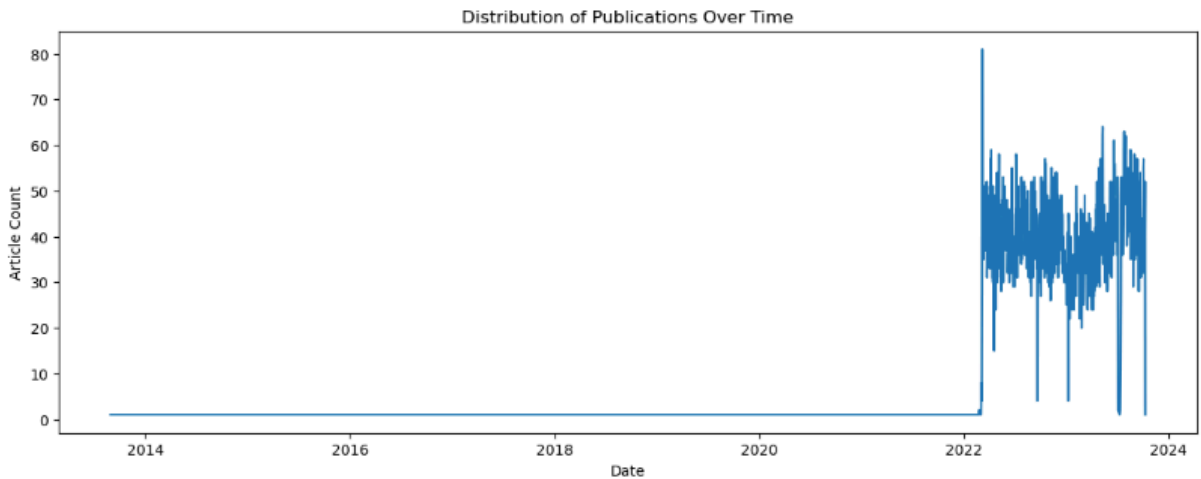

**Figure 2.** BBC news dataset

These preprocessing steps improve the reliability of lexicon-based sentiment scoring by reducing sparsity and token mismatch errors [14].

**Lexicon-Driven Sentiment Analysis.** A multi-lexicon sentiment analysis module is adopted to capture complementary aspects of sentiment signals [13-14]. Four widely used lexicons are employed: VADER [4], optimized for short, and opinionated text with rule-based heuristics [15], AFINN, providing integer-scaled polarity values from −5 to +5 [16], TextBlob, offering polarity via pattern-based subjectivity lexicons [17], and SentiWordNet, assigning sentiment weights to synsets derived from WordNet [18]. Each article is processed independently by all four lexicons [15-18]. For each system, a numerical polarity score is produced. These raw lexicon outputs form a *four-dimensional lexicon-score vector*, reflecting heterogeneous sentiment interpretations of the same article.

**Hybrid Sentiment Modelling.** To synthesise the lexicon outputs into a unified sentiment judgment, a hybrid classification model is constructed using multinomial logistic regression [19]. The lexicon-score vector serves as the input feature vector, and the model outputs a probabilistic sentiment distribution over three classes:

$$s=[P(Negative),P(Neutral),P(Positive)] \qquad (1)$$

The hybrid model mitigates inconsistencies inherent in individual lexicons and benefits from the complementary strengths of rule-based and statistical sentiment estimators. This produces a stable and interpretable sentiment encoding suitable for sequential decision-making.

**Sentiment State Representation.** The probabilistic sentiment vector *s* is treated as the state representation for the reinforcement learning agent [20]. Reinforcement learning (RL) in this context refers to a machine learning paradigm where an agent learns optimal news classification and recommendation strategies by interacting with the environment and receiving feedback [20]. In Sentiment-Guided Q-Learning, article-level sentiment scores shape the reward signal, enabling the agent to adaptively choose actions such as classifying or recommending articles that maximise long-term user engagement and predictive accuracy. Unlike discrete sentiment labels, the probability distribution conveys uncertainty and subtle emotional gradients between articles. This richer representation allows the agent to learn more nuanced policies that reflect user-sentiment alignment rather than rigid category matching.

**Reinforcement Learning Agent.** A Q-learning algorithm is adopted as the recommendation controller. At each interaction step: The agent observes the sentiment-state vector *s*. It selects one of three possible actions: Recommend Positive, Recommend Neutral, or Recommend Negative. A reward signal is computed based on simulated or historical user engagement (e.g., click, dwell time, or preference match). The Q-table is updated according to:

$$Q(s,a) \leftarrow Q(s,a) + \alpha[r + \gamma \max{}'Q(s',a') - Q(s,a)] \qquad (2)$$

where $\alpha$ is the learning rate and $\gamma$ is the discount factor. This iterative update process allows the agent to discover sentiment-aligned recommendation patterns that maximise expected future engagement [20]. Through repeated interaction with the environment, the Q-learning agent converges toward an optimal sentiment-aware recommendation policy [21]. This policy maps probabilistic sentiment states to action preferences, enabling dynamic adaptation to evolving user affective patterns and news content sentiment distributions. The final system shown in Figure 3 thus functions as an adaptive recommender capable of aligning recommendations with user sentiment tendencies while continuously improving based on experience.

**Evaluation.** After the reinforcement learning agent converges to an optimal policy, system performance is evaluated using standard classification and recommendation quality metrics [20]. A confusion matrix is constructed to compare the predicted sentiment-based recommendations (positive, neutral, negative) with the ground-truth sentiment labels derived from the hybrid sentiment model [4].

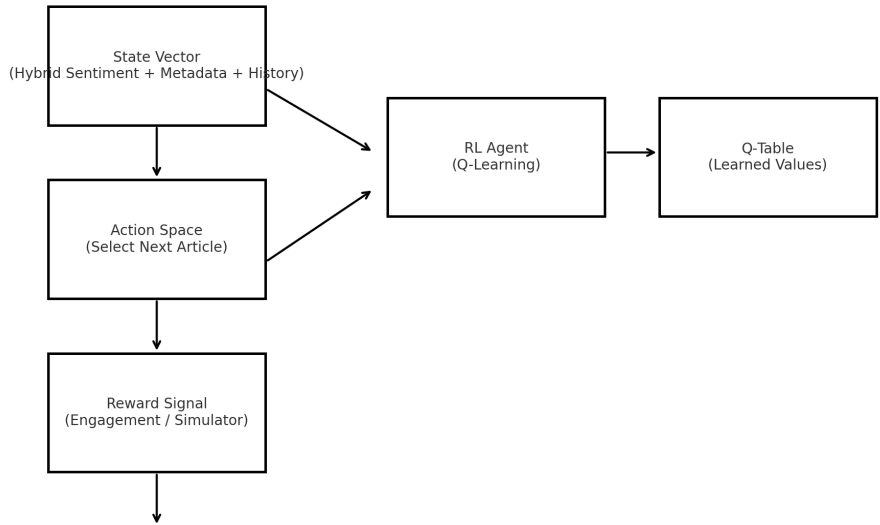

**Figure 3.** Optimal policy learning

From this confusion matrix, precision, recall, and F1-score are computed for each sentiment class:

- **Precision** measures the proportion of recommended articles whose sentiment category matches the ground truth.

- **Recall** quantifies the system's ability to correctly retrieve all items of a given sentiment class.

- **F1-score**, the harmonic mean of precision and recall, provides a balanced measure of accuracy under class imbalance.

These evaluation metrics collectively assess how effectively the optimal Q-learning policy aligns sentiment-based recommendations with the true sentiment distribution of the dataset, thereby indicating the system's recommendation reliability and affective consistency [20-21].

## IV. Results

Figure 4 presents a t-SNE projection of the BBC News dataset, where high-dimensional text embeddings are reduced to two dimensions to visually examine topical structure. Each point represents an individual news article, and colors indicate manually assigned topic groups such as Ukraine War, COVID-19, Sports, and UK Politics. The dense grey region labeled "Other" reflects the large proportion of general-interest articles that do not fall into the highlighted categories. Although t-SNE does not enforce strict cluster separability, several topical tendencies emerge: Ukraine War articles (red) form noticeable localized concentrations, indicating strong semantic similarity; COVID-19 articles (blue) appear moderately dispersed with some compact pockets, reflecting overlapping vocabulary with other public-health or policy-related news; and Sports articles (green) remain relatively sparse but still exhibit localized grouping (Figure 4).

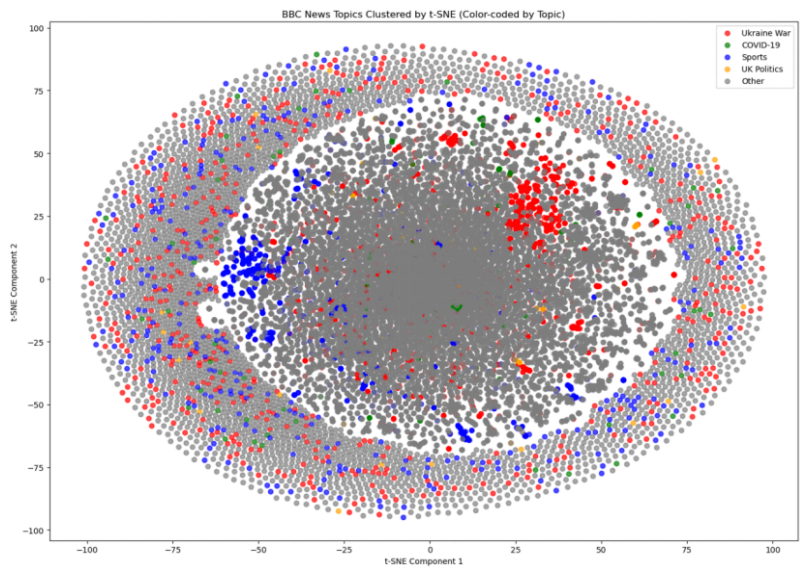

**Figure 4.** t-SNE projection of the BBC news dataset

This visualization illustrates the heterogeneity of the dataset and highlights that certain topics produce tighter semantic clusters, whereas others diffuse across the embedding space. This supports the use of hybrid sentiment modelling and reinforcement learning, as the dataset's natural topical dispersion implies that both sentiment cues and contextual structure are essential for reliable recommendation. Figure 5 illustrates the distribution of the most informative terms across a subset of top-ranked BBC News articles, providing a compact view of how distinguishing vocabulary varies between documents.

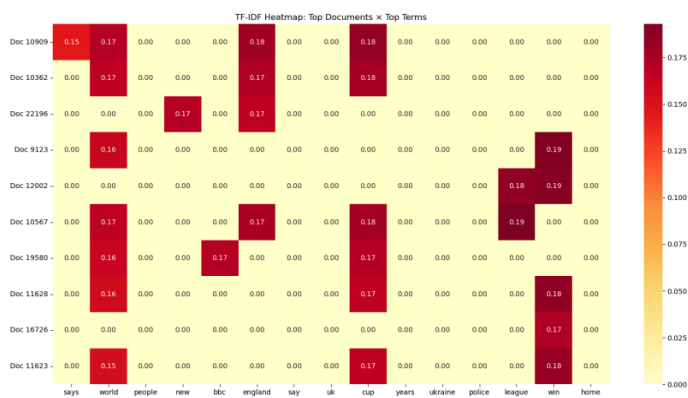

**Figure 5.** The TF-IDF heatmap

Each row represents a document and each column corresponds to a high-value term, with darker red intensities indicating higher TF-IDF scores and thus stronger term relevance within that specific article. The sparsity of the matrix where most cells remain near zero reflects the discriminatory nature of TF-IDF, which suppresses common vocabulary while amplifying topic-defining words such as *cup*, *league*, *england* (sports), *police* (crime), and *Ukraine* (international affairs). This distribution reveals clear topical separation among documents, where each article exhibits high TF-IDF weights only for a small set of semantically coherent terms. Consequently, the heatmap supports exploratory data analysis by highlighting document–term relationships, facilitating topic understanding, category differentiation, and the identification of patterns that inform downstream tasks such as clustering, sentiment classification, and reinforcement-learning–based recommendation. For modelling and comparative evaluation, the analysis focused on four primary tools: *TextBlob, VADER, SentiWordNet, and AFINN*. These tools are used as the main techniques for determining the sentiment of the news description and were applied independently to the BBC dataset to generate baseline sentiment scores. The previous steps allowed the production of the `bbc_news_processed.csv` dataset which include `clean_text`, tokens and `indexed_tokens` to the existing fields. Figure 6 shows the sentiment scores and the label counts of each sentiment lexicon. The AFFINN scores show both the sentiment scores applied on raw text as well as the normalized text since normalization helps with token average to allow better comparison with other tools.

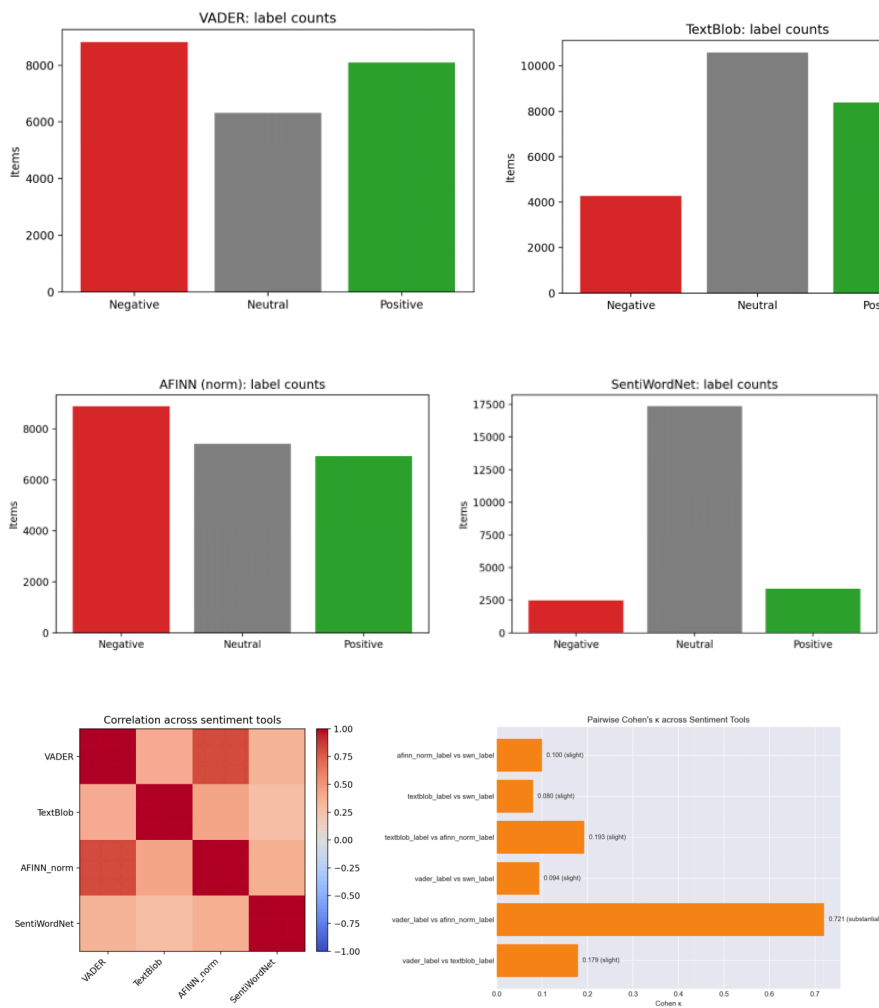

**Figure 6.** Lexicon's sentiment analysis

**Development of the Hybrid Sentiment Analysis Model**

The hybrid sentiment analysis model was developed by integrating the polarity outputs of four established lexicon-based tools (VADER, AFINN, SentiWordNet, and TextBlob) into a unified predictive framework. A cleaned BBC News corpus (bbc_news_with_sentiment.csv) was enriched with both numerical sentiment scores (vader_compound, textblob_polarity, afinn_norm, swn_score) and corresponding rule-based categorical labels for a three-class scheme (Negative, Neutral, Positive) (Figure 6). Because the dataset lacked human-annotated sentiment labels, a weak-supervision strategy was employed by generating a consensus target through majority voting across the four tools, using only non-tied cases for training and evaluation to ensure label reliability. A multinomial logistic regression classifier was then trained using a pipeline consisting of StandardScaler and class-weighted LogisticRegression (max_iter=200), operating on a four-dimensional standardized feature vector and optimized with an 80/20 stratified split. This hybrid model employs a weighted, rule-informed consensus mechanism that stabilises sentiment predictions by combining VADER's intensity-aware scoring with the linguistic nuance of TextBlob and the lexicon-driven polarity structure of AFINN and SentiWordNet. Performance evaluation on the held-out test set demonstrates that the hybrid model surpasses the individual tools like VADER, TextBlob, and SentiWordNet in both accuracy and macro-F1, confirming that the fused representation yields more robust sentiment classification for subtle and context-dependent news content (Figure 7 and Table 1).

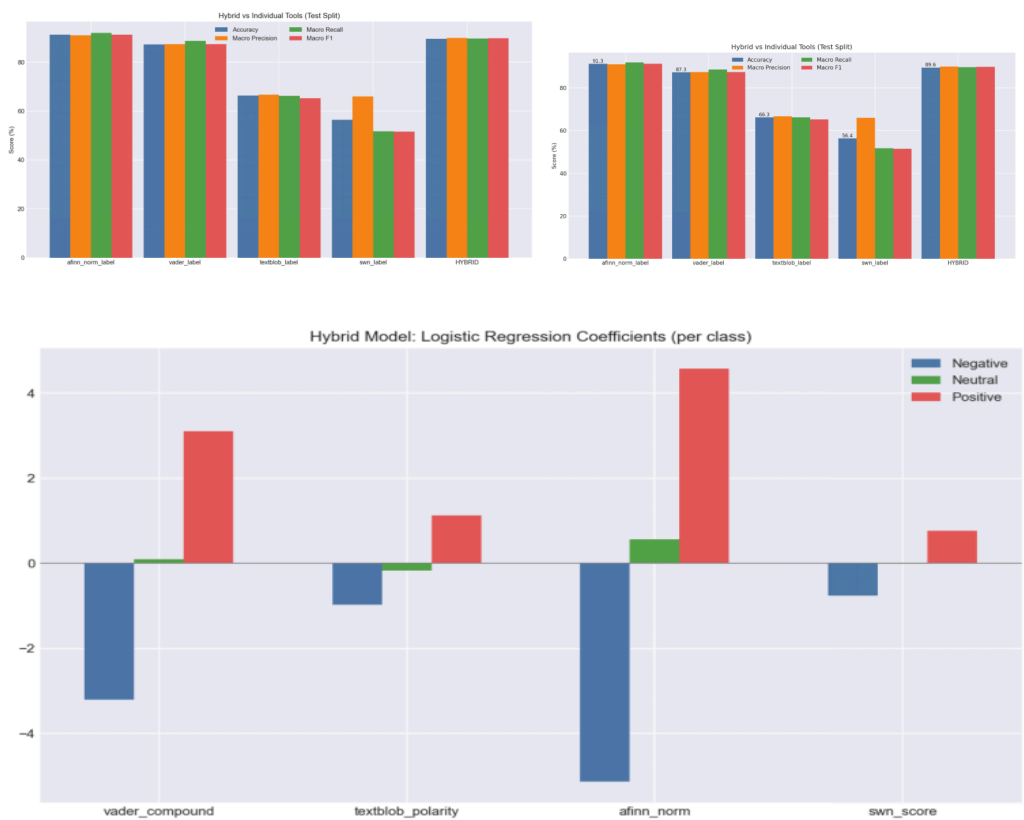

**Figure 7.** Hybrid ensemble analysis

**Reinforcement Learning for Sentiment-Aware News Recommendation**

Reinforcement Learning (RL) was employed to develop a personalised news recommendation system driven by the hybrid sentiment model's probabilistic outputs. In this framework, the RL state encodes the three-way sentiment probabilities produced by the multinomial logistic regression model, representing Negative, Neutral, and Positive polarity supplemented with article metadata and user interaction history. The agent's actions correspond to selecting the next article to recommend, while rewards are obtained either from real user engagement signals or, in the absence of such data, a simulator-based proxy designed for prototyping.

**Table 1.** Comparative analysis of experimental models

| Model / Tool | Accuracy | Macro-F1 Score | Notes |
|---|---|---|---|
| **Hybrid** | 0.896 | 0.898 | Combines all four tools as input features |
| **AFINN** | 0.913 | 0.913 | Best performing baseline |
| **VADER** | 0.873 | 0.874 | Strong baseline performance |
| **TextBlob** | 0.663 | 0.653 | Moderate performance |
| **SentiWordNet** | 0.564 | 0.516 | Lowest performing baseline |

Two scripts were implemented: one suited for deployment in real environments requiring user activity logs, and a second that uses a simulated environment to demonstrate the system's operation. The RL agent was trained using Q-learning, which iteratively updates action value estimates to maximise long-term reward by aligning recommendations with inferred user sentiment preferences. The resulting Q-table summarizes the learned behaviour by capturing the expected cumulative reward associated with recommending Negative, Neutral, or Positive content from each sentiment state (Table 2). This integration of hybrid sentiment as both a contextual descriptor and a behavioural signal enables the agent to learn personalised, sentiment-congruent recommendation strategies even in low-interaction or weak-supervision settings.

| State (Sentiment) | Recommend Negative | Recommend Neutral | Recommend Positive |
|---|---|---|---|
| Negative | 8.748895 | 8.730850 | 8.741671 |
| Neutral | 8.603612 | 8.634936 | 8.635025 |
| Positive | 12.509972 | 12.495487 | 12.532446 |

**Table 2.** Q-table learned by the RL Agent

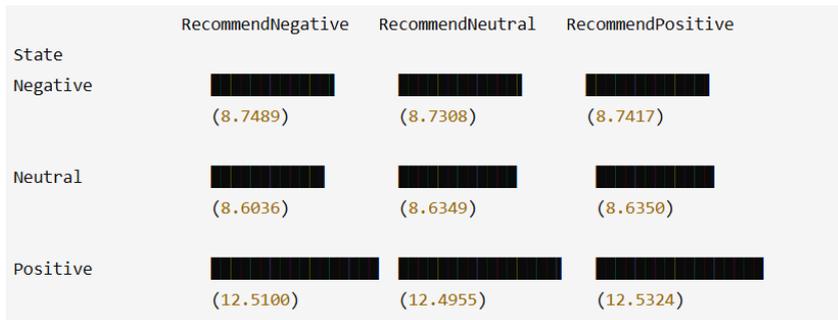

**Figure 8.** Q-table heatmap

## V. Discussions

The Q-table analysis shows that the reinforcement learning agent consistently assigns the highest long-term rewards to recommending Positive news articles, regardless of the current sentiment state (Figure 8). While Negative and Neutral states yield moderate Q-values (≈8.6–8.75), the Positive state produces substantially higher values (≈12.5), indicating a strong asymmetry that favours positivity (Figure 8). This pattern suggests that the agent has learned that positive content maximises cumulative reward either due to simulated engagement signals or inherent behavioural tendencies toward uplifting news. As a result, the learned optimal policy is uniform across all states: RecommendPositive, reflecting a sentiment-aligned recommendation strategy that prioritises emotionally uplifting content for sustained engagement.

| Article | State | Selected Action | P(Neg) | P(Neu) | P(Pos) |
|---|---|---|---|---|---|
| 5169 | Neutral | RecommendPositive | 0.0031 | 0.9029 | 0.0940 |
| 6788 | Positive | RecommendPositive | 0.0002 | 0.2659 | 0.7339 |
| 20853 | Negative | RecommendNegative | 0.9992 | 0.0008 | ~0.0000 |
| 5233 | Neutral | RecommendPositive | 0.0072 | 0.9353 | 0.0575 |
| 22584 | Positive | RecommendPositive | ~0.0000 | 0.0000 | 0.9999968 |
| 20294 | Positive | RecommendPositive | ~0.0000 | 0.0000 | 0.9999972 |
| 4288 | Positive | RecommendPositive | ~0.0000 | 0.0000 | 0.9999941 |
| 6737 | Positive | RecommendPositive | 0.00003 | 0.2244 | 0.7756 |
| 17469 | Positive | RecommendPositive | ~0.0000 | 0.0003 | 0.9997411 |
| 6243 | Negative | RecommendNegative | 0.6795 | 0.3204 | 0.0001 |

**Table 3.** Sentiment recommendations

Table 3 highlights that items with high positive probabilities (P(Pos) > 0.75) consistently trigger RecommendPositive, indicating stable alignment between sentiment and recommended action. Examples with dominant negative probabilities (e.g., 0.9992 or 0.6795) are correctly mapped to RecommendNegative. Interestingly, entries with neutral-dominant probabilities often result in RecommendPositive, reflecting the system's bias toward positive recommendations under the placeholder training reward.

## VI. Conclusion

This study presents an optimal sentiment-based news recommendation framework that integrates exploratory data analysis (EDA), multiple sentiment lexicons, a hybrid logistic regression model, and reinforcement learning. Analyzing the BBC news dataset, the research aimed to classify sentiment accurately, evaluate lexicon-derived features, and explore how adaptive learning can enhance recommendation strategies. The hybrid sentiment model outperformed individual lexicons by combining complementary polarity signals, yielding more consistent and context-sensitive predictions. Reinforcement learning further leveraged these sentiment states to personalise recommendations, adapting policies based on reward signals and producing interpretable, sentiment-driven actions. The study demonstrates that combining hybrid sentiment modelling with reinforcement learning offers a scalable, interpretable, and effective framework for personalised news recommendation, laying the groundwork for future improvements using richer contextual features and real user engagement data.